\documentclass[slac_one]{revtex4}
\usepackage{graphicx}
\usepackage{fancyhdr}
\renewcommand{\headrulewidth}{0pt}
\renewcommand{\footrulewidth}{0pt}

\begin{document}

\title{Searches at LHC Beyond the Standard Model} %% Paper title goes here

\author{Sebastiano Albergo on behalf of the ATLAS and CMS 
Collaborations\footnote{Presented at the  HEP MAD-07 Conference } }
\affiliation{University of Catania and INFN Catania, Italy}
\begin{abstract}
The discovery potentials of ATLAS and
 CMS experiments at the Large Hadron Collider (LHC) for Supersimmetry (SUSY), 
Extra Dimensions (ED), new Gauge Bosons and R-Hadrons are discussed.  Beyond Standard-Model (BSM) searches 
at LHC  require a 
detailed understanding of the detector performance, reconstruction algorithms and
triggering. Precision measurements of Standard Model (SM) processes are also  mandatory to acquire the necessary 
knowledge of SM background. Both ATLAS and CMS efforts are hence addressed to determine the best 
calibration candles and to design a realistic  plan  for the initial period of  data taking.  
\end{abstract}

\maketitle

\thispagestyle{fancy}

\section{Introduction}

ATLAS~\cite{atlas} and CMS~\cite{cms}  are the two general purpose detectors which are being installed at
 LHC 
for new physics studies in the  TeV Mass range. 
One of the main goals of ATLAS and CMS is to search for  Beyond Standard Model (BSM) signals.
The Standard Model~\cite{SM} 
provides a successful description of the existing data but it is 
considered as a low-energy effective theory with some intrinsic deficiencies. 
The most appealing extension of the SM is SUSY~\cite{SUSY}. It proposes an elegant solution to the
"hierarchy" problem~\cite{SUSY} and provides a good candidate for the dark
matter in cosmology, though at the expense of additional parameters of the theory. 

Much interest  has   been produced by other BSM models which hypotesize 
more than four space-time dimensions. Some of these Extra Dimension models~\cite{add,rs,ed,ued} lead to a gravity 
mass scale in the TeV range, hence predicting relevant signatures in collider particle physics.

Only some  of the BSM signatures which will be addressed at LHC are discussed here. 
Section 2 
is devoted to some relevant topics  in detector aligment and calibration;  Section 3
is devoted to  SUSY searches; Section 4 reports some Extra-Dimension  discovery potentials  and the  Z'  sensitivity 
of ATLAS and CMS in the dimuon decay channel.

\section{Detector aligment and calibration}
The development of alignment, calibration and commissioning procedures is of primary importance to achieve  a good understanding of the detector. For this reason, both the Collaborations are strongly involved in the study of suitable start-up strategies. In this Section, the CMS alignment method is discussed. Results from calibration studies ($t\overline{t}$  "candle") performed by ATLAS are also discussed.
Both items well represent the many  efforts performed by ATLAS and CMS to face the challenging initial phase of 
data taking. 

\subsection{CMS alignment}
The alignment uncertainties of the CMS Tracker and Muon detectors  affect the performances of 
the track pattern recognition and reconstruction.
Since the detector alignment  is expected to progressively improve with integrated luminosity, different
scenarios are usually considered. Here are summarized the two scenarios  concerned with data taking:

\begin{itemize}

\item{} First Data Scenario : corresponding to the very early stage of data taking, it should be reached
before accumulating $100 pb^{-1}$. It assumes 1 mm and
0.2 mrad of relative positioning precision between the Tracker and Muon System. Muon Chambers are located
within the Muon System to 1 mm and 0.25 mrad precision. The Tracker structure and modules relative
misalignment ranges from 3 to 13 $\mu m$ for the Pixel Detector and from 50 to 300 $\mu m$ for the Silicon Strip
Detector (SSD).

\item{} Long Term Scenario : corresponding to the situation of optimal alignment performance, it is expected 
after collecting about $1 fb^{-1}$. In this scenario
the Tracker to Muon System relative misalignment will be of 200 $\mu m$ and 50 mrad, while the SSD
precision is improved by a factor 10 with respect to the previous scenario.
\end{itemize}
The integrated luminosity estimated for each alignment scenario are
related to the number of tracks originated from $ Z\rightarrow  \mu\mu$ and $W \rightarrow \mu\nu$ decays, 
 which are used for track based alignment.

\begin{figure}[h]
\vfill \begin{minipage}[t]{.49\linewidth}
\begin{center}
\includegraphics[width=0.95\linewidth]{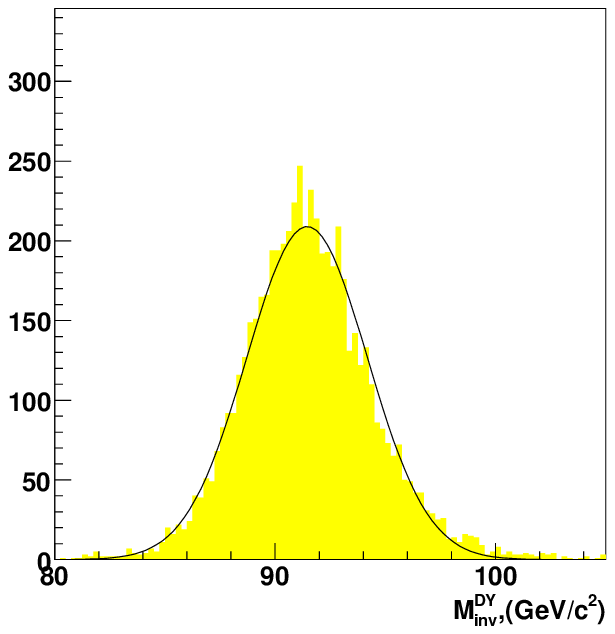}
\end{center}
\caption{  $ Z\rightarrow  \mu\mu$ mass peak at ``First Data Scenario''. $\sigma_M/M = 0.0226$.}
\label{mumu}
\end{minipage}\hfill
\begin{minipage}[t]{.49\linewidth}
\begin{center}
\includegraphics[width=0.95\linewidth]{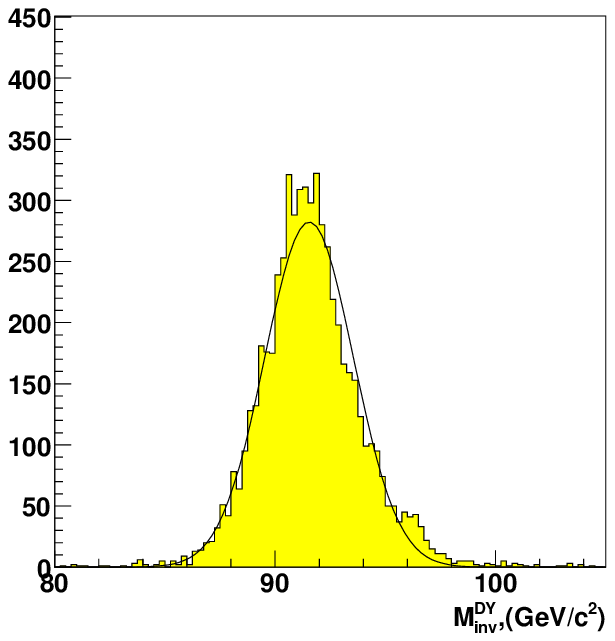}
\end{center}
\caption{$Z\rightarrow  \mu\mu$ mass peak at ``Long Term Scenario''. In this
case $\sigma_M/M = 0.0118$, which is comparable to the value 0.0106  obtained
 in the "ideal alignment" (i.e. no misalignment) case~\cite{belotelov}.}
\label{mumulong}
\end{minipage} 
\end{figure}

 Fig.~\ref{mumu} and  Fig.~\ref{mumulong}
reports the invariant  mass distribution of  $ Z\rightarrow  \mu\mu$  events for   First Data scenario and  Long Term scenario respectively.
The study~\cite{belotelov} is done using simulated events produced by OSCAR~\cite{oscar}, a Monte Carlo 
program based on GEANT4
which fully simulates the passage of particles through the detector taking into account a detailed description of detector
geometry, materials and magnetic field. Digitization of signal generated by simulated particles is  
 done within the framework of the ORCA~\cite{orca} package, which is also used for the event reconstruction.

\subsection{$t\overline{t}$ calibration candle}
The top pair production process is very valuable for the in-situ calibration of the ATLAS and CMS detectors during
the first period of data taking. The top quark decays for nearly $100\%$ to  W$+$b. The  lepton-plus-jet channel,
 $t\overline{t}\rightarrow WWb\overline{b}\rightarrow (l\nu)(jj)(b\overline{b} )$ has
 a branching ratio of $29.6\%$. The large cross section and the large signal to background  (S/B) ratio for the lepton+jets  
 $t\overline{t}$-decay channel, allow to select high purity samples with good statistics with small integrated luminosity. 
These first collected top data samples will be used  to
understand some important performance issues like: detector uniformity, absolute energy scale calibration,
missing transverse energy  calibration, b-tagging, etc.
In addition, with these top samples the Monte Carlo can be tuned and the  background understanding 
can be improved.

\begin{figure}[h]
\vfill \begin{minipage}[t]{.49\linewidth}
\begin{center}
\includegraphics[width=0.95\linewidth]{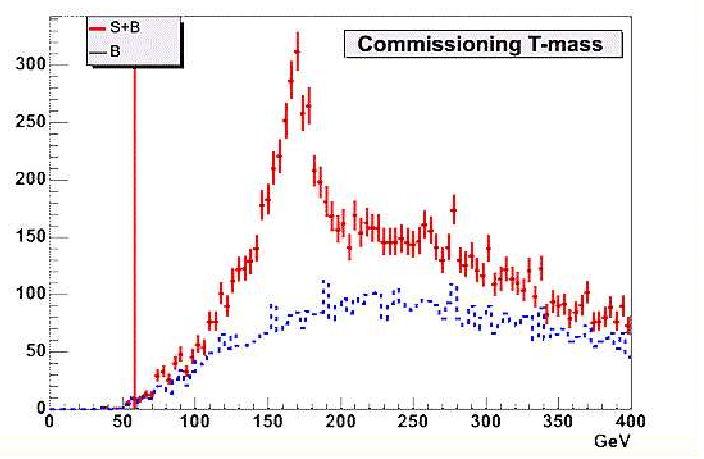}
\end{center}
\caption{Top reconstructed mass with 150 pb$^{-1}$ of integrated luminosity in ATLAS.}
\label{ttbar}
\end{minipage}\hfill
\begin{minipage}[t]{.49\linewidth}
\begin{center}
\includegraphics[width=0.95\linewidth]{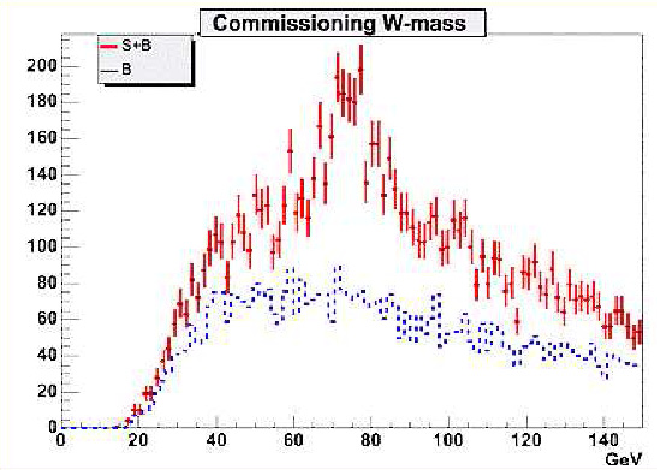}
\end{center}
\caption{W reconstructed mass with 150 pb$^{-1}$ of  integrated luminosity in ATLAS. }
\label{ww}
\end{minipage} 
\end{figure}

In the worst initial scenario of absence of
b-tagging the largest irreducible contribution to the background
originates from W+4 jet events, where the
W-boson decays leptonically and produces an isolated lepton and
missing transverse energy ($E\!\!\!\!/_{\,{\mathrm T}}$), and the four jets survive the selection criteria.
In this scenario the  $t\overline{t}$ candidate events can be selected~\cite{ttbar}   requiring:

\begin{itemize}
\item{} one isolated lepton (electron or muon)with $P_T > 20$ GeV/c

\item{}   $E\!\!\!\!/_{\,{\mathrm T}}> 20$  GeV

\item{} 4 reconstructed-jets with constrained cone sizes, each with $-2.5<\eta< 2.5$ and $P_T > 40$ GeV/c.
\end{itemize}
In Fig.~\ref{ttbar}
the  resulting top mass peak   for 150 pb$^{-1}$ of integrated luminosity in ATLAS is shown~\cite{ttbar} .
An effective method for absolute energy scale calibration relies on the W mass peak reconstruction, by taking
advantage of the well known experimental value of the W mass. 
The W candidates can be selected from the top event set 
by requiring the two jets with highest $P_T$. Plotting the two-jet invariant mass distribution a clear W peak is 
resolved (see  
Fig.~\ref{ww}).

\section{Supersimmetry}

The LHC potentials to discover SUSY are mainly analysed in the more constrained 
framework of the Minimal Supergravity (mSUGRA)~\cite{msugra} model with only 5 free
parameters: the common scalar ($m_0$) and fermion ($m_{1/2}$) masses, the trilinear coupling 
($A_0$) and the Higgs sector parameters ($tan\beta$, $sgn\mu$) at the Grand Unification (GUT) scale. 
Assuming R-parity, new supersymmetric particles are produced in pairs and 
the lightest one, namely the Lightest Supersimmetric Particle (LSP), is stable and neutral.
\subsection{Inclusive signatures}
 At the LHC, the SUSY production is 
dominated by strongly interacting squarks and gluinos, which have long decay 
cascades with the jet emission. The cascade ends with the LSP, which is not detected. 
Therefore, the most generic SUSY signature is a multi-jet 
with large  $E\!\!\!\!/_{\,{\mathrm T}}$ final state. 
Leptons produced in decays of charginos or neutralinos 
can also be present, hence also final states with (n$\ge 1$ 
leptons)+jets+$E\!\!\!\!/_{\,{\mathrm T}}$ are considered. 
The main backgrounds are QCD and ${\mathrm t\overline{\mathrm t}}$, W and Z with QCD-jet 
associated production processes. Finally, diboson production,
such as WW+jets, WZ+jets, and ZZ+jets, also contribute as sources of background. 
Considering signatures with at least one lepton, substantially reduces the QCD background. 

Due to the very high QCD production cross section, the main background for 
the jets+$E\!\!\!\!/_{\,{\mathrm T}}$ channel is dominated by QCD events with  
large missing transverse energy resulting from jet mismeasurements 
and detector resolution effects. Topological variables are used to reduce 
as much as possible the QCD background, in particular the angular correlation between 
the first two jets (ordered in $E_{\mathrm T}$) and the $E\!\!\!\!/_{\,{\mathrm T}}$ direction 
is used.

The inclusive  SUSY searches employ the following strategy~\cite{tdr}. 
First, experimental signatures are studied for a limited number of test points 
of mSUGRA parameter space
using the full  simulation and reconstruction  software. 

\begin{figure}[h]
\vfill \begin{minipage}[t]{.49\linewidth}
\begin{center}
\includegraphics[width=0.60\linewidth]{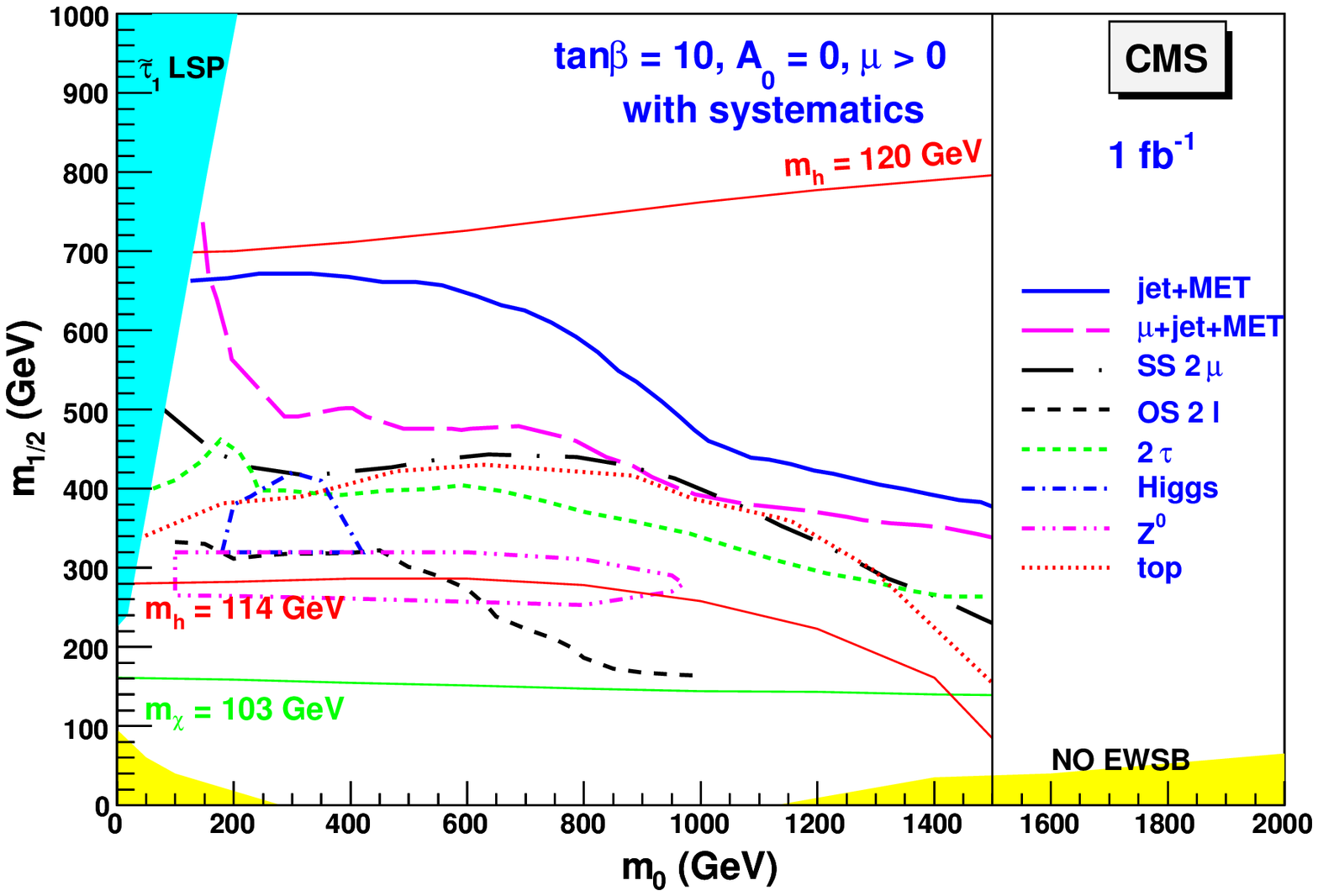}
\end{center}
\caption{The mSUGRA discovery reach 
in the ($m_0$--$m_{1/2}$) plane for fixed $A_0=0$, $tan\beta=10$, $\mu>0$ 
for 1 fb$^{-1}$ of collected data for several search
strategies studied by CMS (including systematic uncertainties) as presented in Ref.~\cite{tdr}.}
\label{fig.1}
\end{minipage}\hfill
\begin{minipage}[t]{.49\linewidth}
\begin{center}
\includegraphics[width=0.65\linewidth]{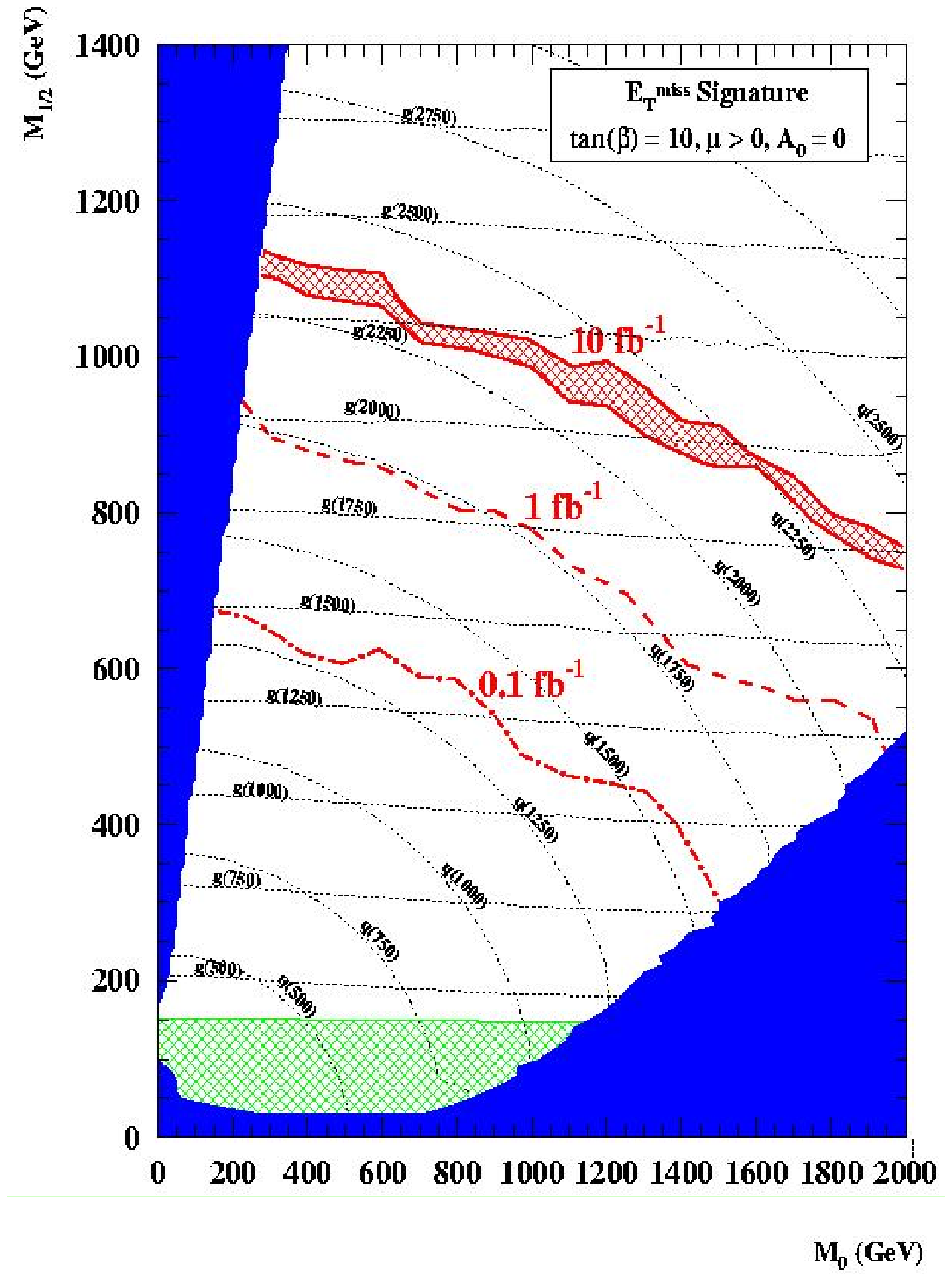}
\end{center}
\caption{The mSUGRA discovery reach of ATLAS for the jets+$E\!\!\!\!/_{\,{\mathrm T}}$ inclusive channel
in the ($m_0$--$m_{1/2}$) plane for fixed $A_0=0$, $tan\beta=10$, $\mu>0$ and
for different values of integrated luminosity.}
\label{susyatlas}
\end{minipage} 
\end{figure}
Next, the results are extended to other points of the parameter space. 
using fast simulation and reconstruction. 
In order to obtain the best signal to 
background (S/B) ratio the SUSY selection cuts are optimised for
each point. 
The expected discovery reach is evaluated for parameter sets 
having at least five standard deviation (5$\sigma$) signal significance. 
As shown in Fig.~\ref{fig.1}, referred to CMS, several different final states  have been considered~\cite{tdr}. 

As can be seen in Fig.~\ref{susyatlas}, where the ATLAS discovery potential 
for the jets+$E\!\!\!\!/_{\,{\mathrm T}}$ inclusive channel is shown, a huge region of the parameter space can be exploited already with low integrated luminosity. 

The multi-jets $+E\!\!\!\!/_{\,{\mathrm T}}$ 
final state  is
expected to be the most sensitive SUSY-search strategy.

Signatures involving at least one lepton are less sensitive, but experimentally cleaner and 
have the advantage of an efficient and well-understood trigger shortly 
after LHC start-up. Topological requirements on the
jets and missing energy are similar to the fully inclusive analysis.

\subsection{Semi-inclusive and esclusive searches}
Exclusive SUSY searches consist in reconstruction of specific decay
channels in order to estimate physical parameters characterising the decay itself. Due to the presence of the invisible LSP, a direct determination of the sparticle masses involved in the decay is not possible at LHC. A well established technique to achieve this goal exploits the presence of kinematic properties (end points and thresholds) 
in the invariant mass distribution of various subsets of particles 
(usually leptons and quarks and their combinations)~\cite{edges}. 
This  kind of analysis 
is able to give strong constraints on mass spectrum and SUSY parameters.

\begin{figure}[h]
\vfill \begin{minipage}[t]{.49\linewidth}
\begin{center}
\includegraphics[width=0.750\linewidth]{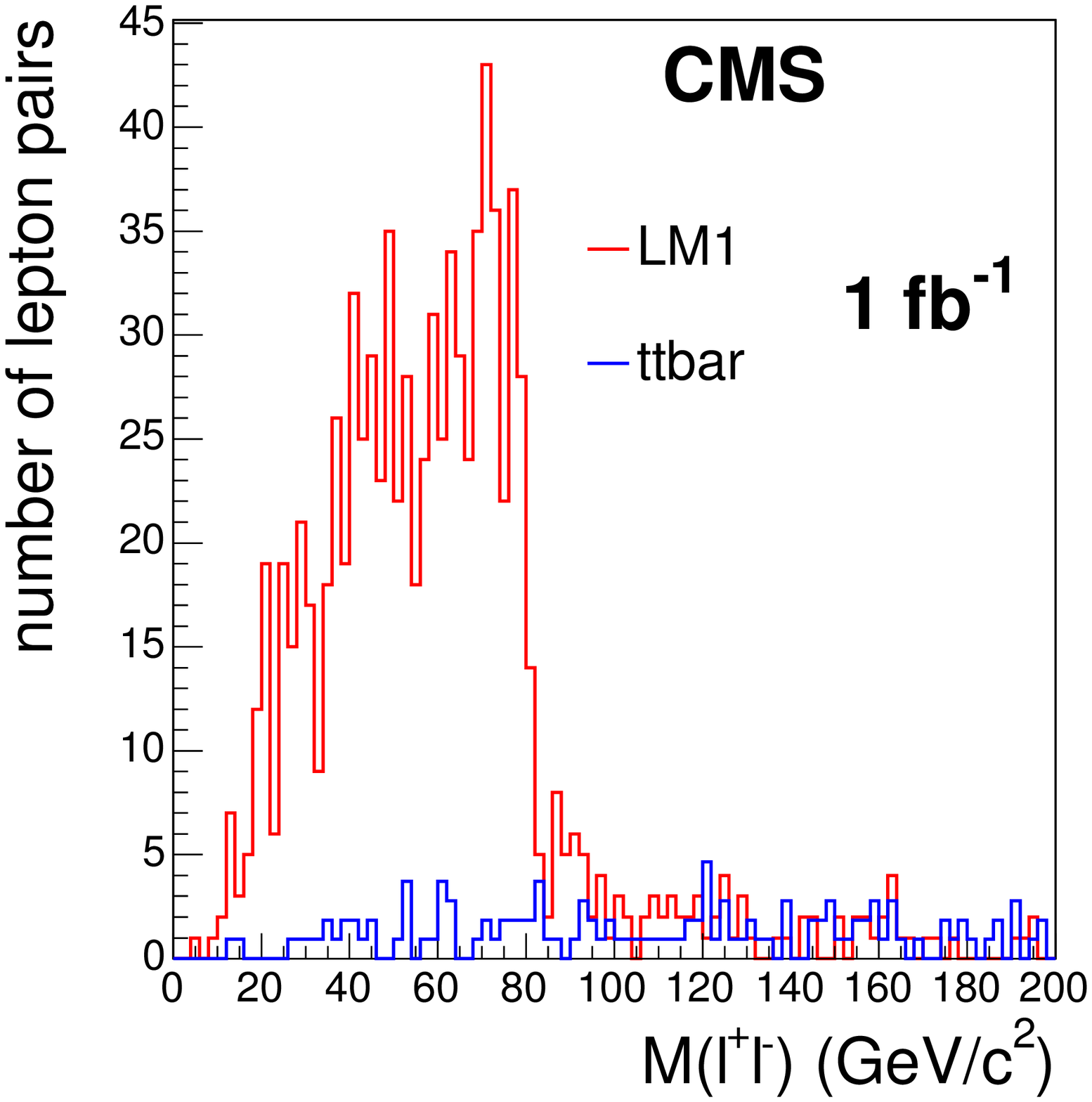}
\end{center}
\caption{Same flavour opposite sign lepton pair distributions
of SUSY and ${\mathrm t\overline{\mathrm t}}$ events for 1 fb$^{-1}$. The peculiar 
triangular edge is clearly visible at point LM1.}
\label{fig.2}
\end{minipage}\hfill
\begin{minipage}[t]{.49\linewidth}
\begin{center}
\includegraphics[width=0.90\linewidth]{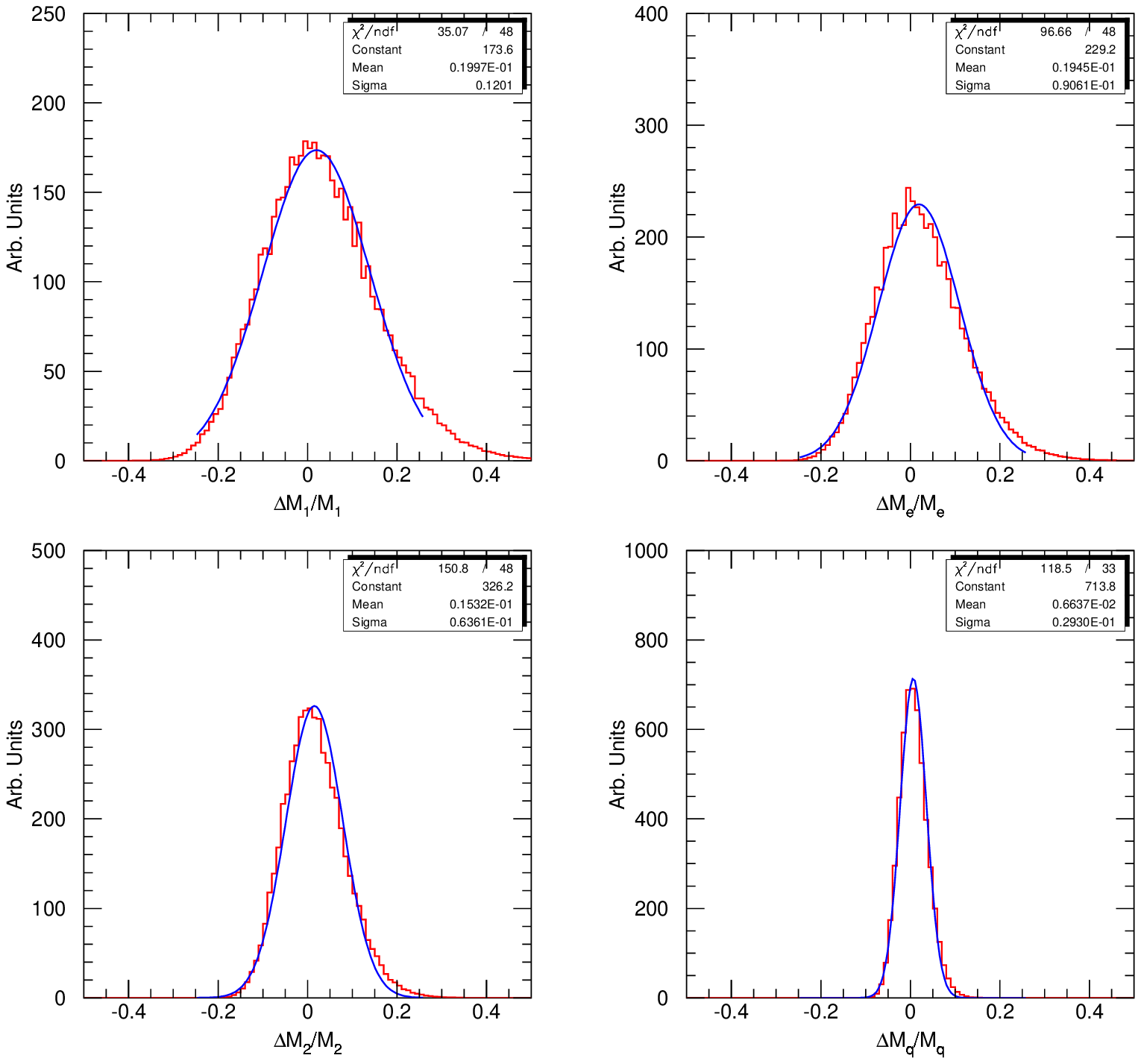}
\end{center}
\caption{Distribution of the   $\tilde{\chi}_1^0$, $\tilde{l_R}$, $ \tilde{\chi}_2^0$
and  $\tilde{q}_L$ invariant masses at 70 fb$^{-1}$ of integrated luminosity. 
The fitted width are $\pm 12\%$, $\pm 9\%$, $\pm 6\%$ and 
$\pm 3\%$ respectively.}
\label{susymass}
\end{minipage} 
\end{figure}

Using similar cuts of the inclusive case  and looking for same flavour opposite sign (SFOS)
dilepton pairs provides a clean SUSY signature. The SFOS dilepton final state is particularly interesting
since leptons (electrons and muons) from the $\tilde{\chi}_2^0$ decay exhibit a peculiar $l^+l^-$
 invariant mass distribution with a sharp edge, as shown for CMS in Fig.~\ref{fig.2} at the
parameter point LM1 ($m_0=60$ Gev/c$^2$, $m_{1/2}=250$ Gev/c$^2$, $tan\beta=10$, $A_0=0$, $\mu>0$) in a full
simulation and recostruction study~\cite{nostranota}. 
The reconstruction of 
the dilepton edge provides information on the sparticles involved in the decay chain and
constitute a powerfull indication of new physics especially in the early data taking period.

In   Ref.~\cite{susymass}  an esclusive study is presented which shows an example of sparticle mass reconstruction
with the ATLAS detector in the three-step decay chain  $\tilde{q}_L\rightarrow \tilde{\chi}_2^0  q\rightarrow  
\tilde{l}lq \rightarrow \tilde{\chi}_1^0 l^+l^-q$. The simulated data   corresponds to about 
70 fb$^{-1}$ of integrated luminosity and are generated in the mSUGRA framework with parameter values 
$m_0=100$GeV/c$^2$, $m_{1/2}=300$GeV/c$^2$,  $A_0=300$GeV, $tan\beta=2$, $sgn\mu=+$. Suitable cuts are adopted
 on  $E\!\!\!\!/_{\,{\mathrm T}}$,  on the number of jets and isolated leptons and on their
$P_T$ and $\eta$ values.  The reconstructed edges of ll, lq and llq invariant mass distribution together with the llq 
threshold are obtained in order to extract, by fitting procedure, the resulting  
 $\tilde{\chi}_1^0$, $\tilde{l_R}$, $ \tilde{\chi}_2^0$
and  $\tilde{q}_L$ mass distributions. In order to properly account  the uncertainties on the four end-point values, a 
huge number of randomly generated sets of    $\tilde{l_R}$, $ \tilde{\chi}_2^0$
and  $\tilde{q}_L$ mass values has been produced. Each set  has been weighted depending on
the dispersion on the kinematically related  $\tilde{\chi}_1^0$ mass values. Resulting mass histograms are shown
in  Fig.~\ref{susymass}.

\subsection{R-Hadrons}
Different SUSY models predict long-lived  sparticles which can hadronize with normal quarks or gluons
forming heavy hadrons, named R-Hadrons. Split Supersymmetry models (Split SUSY~\cite{giudice})  predict 
a long lived gluino. 
The existence of  a long lived stop is hypotesized in some  SUSY-5D~\cite{barbieri}. 
The new long-lived hadrons are expected to have lifetime high enough to cross the detector and a mass
higher then $\sim 100$ GeV/c$^2$. R-hadrons can flip their electric charge in hadronic interactions
when crossing the detector. Being massive these particles are slow in spite of a high
momentum and their specific energy loss  is much higher, at a given momentum,
than  ordinary particles. 

\begin{figure}[h]
\begin{center}
\includegraphics[width=0.85\linewidth]{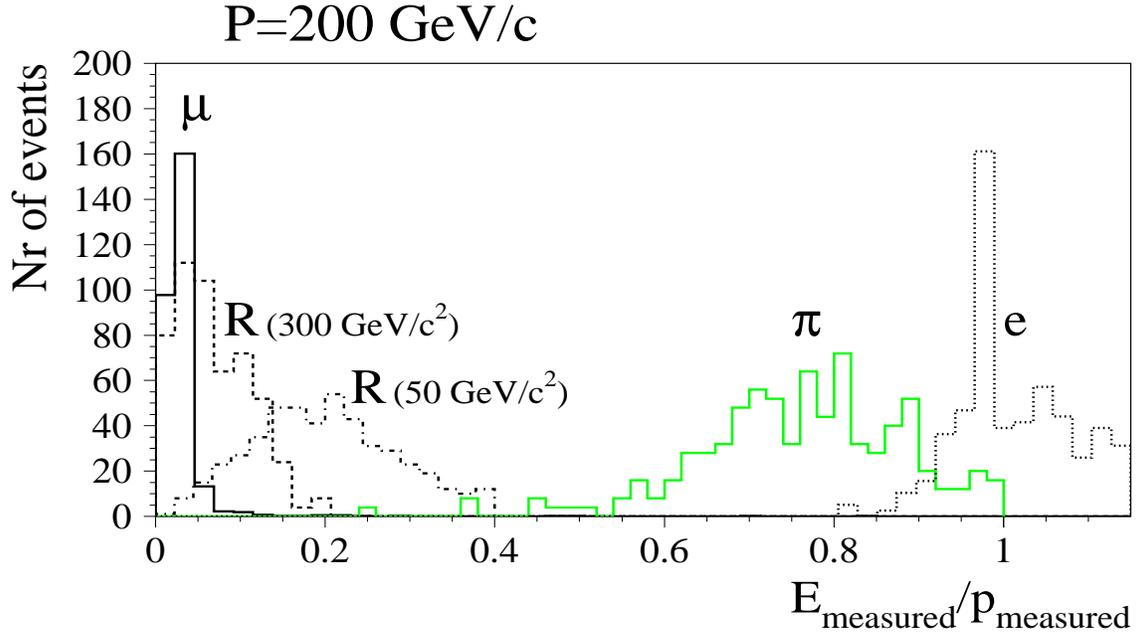}
\end{center}
\caption{The ratio E/p for R-hadrons, muons, pions and electrons $\eta = 0.1$. Singly
charged R-mesons, muons, pions and electrons are generated and reconstructed in fast simulation.}
\label{rhadrons}
\end{figure}

It has been shown  with a fast simulation study~\cite{rhadrons} 
that calorimeter measurements could be used to identify R-hadrons. In Fig.~\ref{rhadrons}
particle distributions vs E/P ratio for different kind of 
incident particles impinging the ATLAS barrel calorimeters are  plotted.

Another ATLAS  study~\cite{hellman}  
performs the recontruction of
 gluino R-hadron masses with a method~\cite{polesello} based on  time of flight measurements, 
 while, together with the time of fligh method, a complementary
 approach~\cite{rizzi}   
is studied in CMS. This approach is based on the specific energy loss measurement inside the CMS 
silicon  tracker. R-hadron velocity is then indirectly obtained by inverting
the Bethe-Bloch formula in the range $0.1<\beta<0.9$.

\section{Other BSM searches}
\subsection{Gravitons}
The production of Kaluza-Klein gravitons in high energy particle collision is predicted
in various extra dimensional scenarious. Presently the most popular
extra dimensional models  in high energy physics are the 
Arkani-Hamed, Dimopoulos and Dvali (ADD)~\cite{add};  
the Randall Sundrum (RS)~\cite{rs};
 the TeV$^{-1}$ Size Extra Dimensions~\cite{ed}; 
and the Universal Extra Dimensions (UED)~\cite{ued}.

The ADD model has  large compactified extra dimensions, in which gravity can propagate, while the
SM  particles are confined to the usual 4-dimensional space-time  brane. In the RS model, the
hierarchy problem is solved by having a single highly curved (warped) extra dimension. 
In this scenario, gravity is localized on
one brane in the extra dimension, while the SM particles are located on another. In the TeV$^{-1}$ Size Extra
Dimensional Model, the SM chiral fermions are confined to a brane or branes, but the SM gauge bosons
(W, Z, $\gamma$ and g) can propagate into the extra dimensions. In UED, all SM fields are allowed to
propagate along EDs. Therefore, each SM particle has Kaluza-Klein  excitations.

\begin{figure}[h]
\vfill \begin{minipage}[t]{.49\linewidth}
\begin{center}
\includegraphics[width=0.65\linewidth]{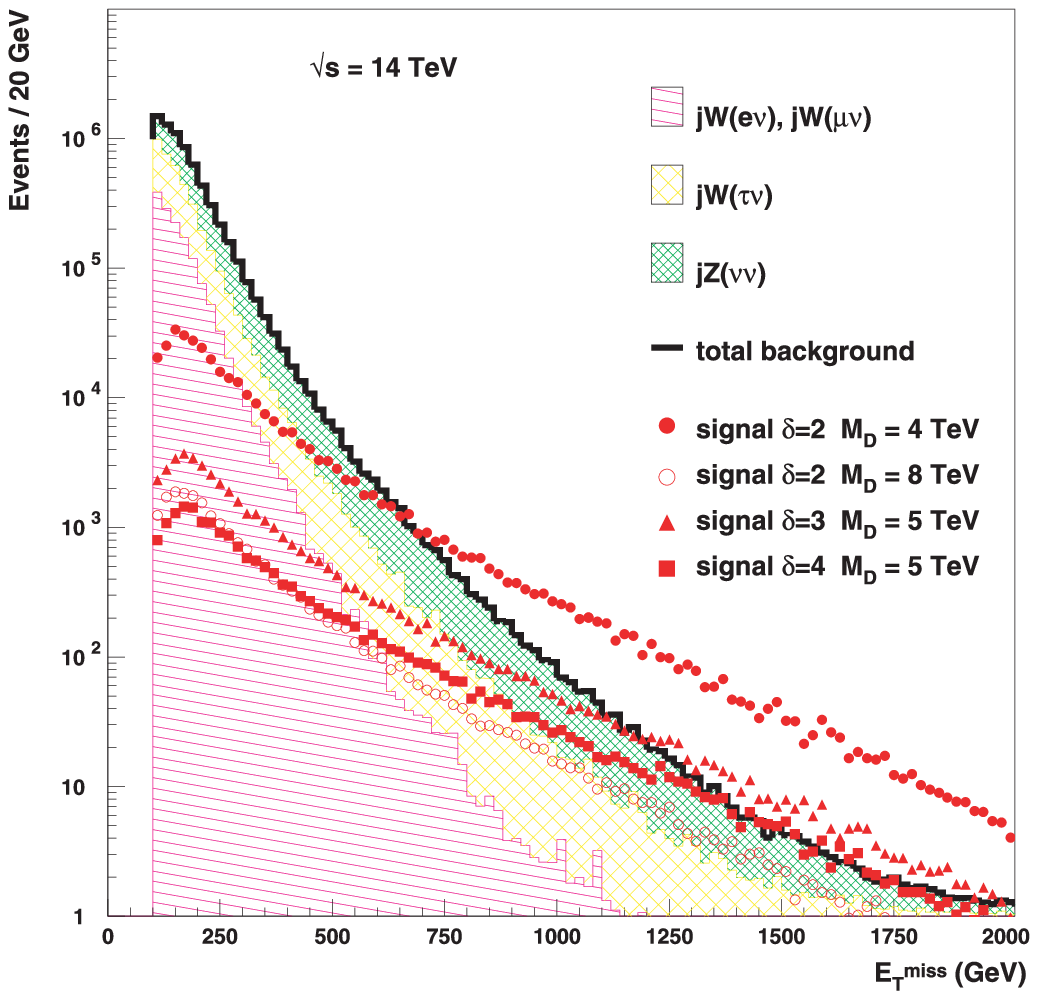}
\end{center}
\caption{ $E\!\!\!\!/_{\,{\mathrm T}}$ distribution  in ATLAS of background events and of ADD-graviton signal events
after data selection, for 100 fb$^{-1}$ of integrated luminosity. 
The contribution of the three main kinds of background is
shown as well as the distribution of the signal for several values of the model parameters.}
\label{atlasadd}
\end{minipage}\hfill
\begin{minipage}[t]{.49\linewidth}
\begin{center}
\includegraphics[width=0.70\linewidth]{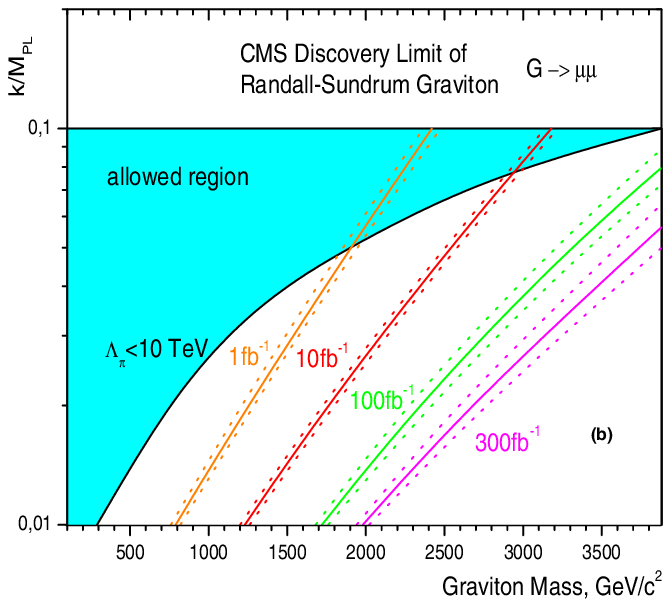}
\end{center}
\caption{CMS Discovery limit for RS graviton  as a function of the model parameter k 
and the graviton mass for various values of integrated luminosity. The left part of each curve is the region 
where significance exceeds 5$\sigma$. The ranges of the expected variations
due to the systematic uncertainties are shown.}
\label{rs}
\end{minipage} 
\end{figure}

All  these models have signatures which are potentially detectable at accelerators. In particular ADD and RS models 
predict  evident signatures based on direct  graviton production.
An emission of a massless graviton in association with a jet is predicted by ADD, so a clear signature
of back-to-back jet+$E\!\!\!\!/_{\,{\mathrm T}}$ is expexted. in which the  $E\!\!\!\!/_{\,{\mathrm T}}$ 
 is related to gravitons emitted into the extra
dimensions, which consequently escape detection.  The dominant backgrounds arise
from processes that can give rise to neutrinos in the final state, namely jet$+ Z\rightarrow \nu\nu$, 
jet$ + W\rightarrow \tau\nu$, jet$ + W\rightarrow \mu\nu$ and jet$ + W\rightarrow e\nu$. 
By  vetoing the events where there is an
isolated lepton within the acceptance of the ATLAS muon or tracking systems the background from the last two 
sources is reduced.
Fig.~\ref{atlasadd}
shows~\cite{atlasadd} the  $E\!\!\!\!/_{\,{\mathrm T}}$ distribution of the backgrounds and
of the signals in ATLAS for several choices of the number of extra dimensions $\delta$ and the model parameter
$M_D$. The signal emerges from the
background at large  $E\!\!\!\!/_{\,{\mathrm T}}$. 
The distributions for the different signals reflect the expected scaling
of the cross section as a function of $M_D^{-\delta-2}$. 

The RS model  graviton  can decay in the dilepton,
dijet or diboson channel.  
 In the dilepton case  the signature  would be a  series of narrow resonances 
in the dilepton  invariant mass distribution.
The dominant background arises from the Drell-Yan  lepton pair production, whereas contributions
from  $t\overline{t}$ and the vector boson pair production (ZZ, WZ, WW) are significantly smaller and are highly
suppressed by selection cuts. The CMS discovery limit~\cite{tdr}  for this channel  is shown in Fig.~\ref{rs}
for 1, 10, 100 and 300 fb$^{-1}$ of integrated luminosity.

\subsection{New Gauge bosons}

Extra Gauge Bosons $Z'$ and $W'$ are predicted by several different models, generally 
belonging to the following classes~\cite{zprime}:
\begin{itemize}
\item{} Superstring inspired and Grand Unification theories (GUT);

\item{} Left-Right Symmetric Models based on the gauge group $SU(3)_C$x$SU(2)_L$x$SU(2)_R$x$U(1)_{B-L}$
  predicting substructures of the known elementary particles; 

\item{} Little Higgs Models. 
\end{itemize}

\begin{figure}[h]
\begin{center}
\includegraphics[width=0.60\linewidth]{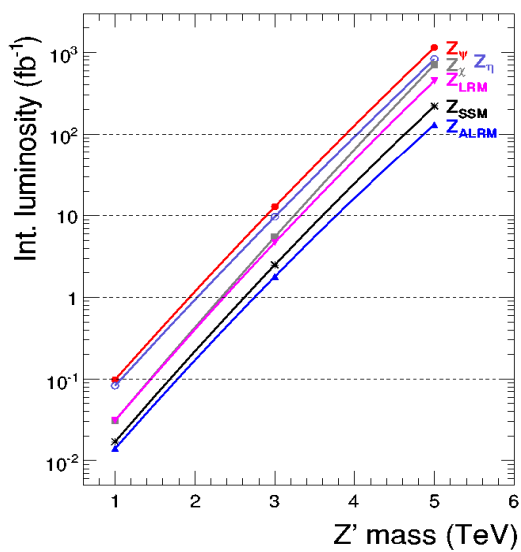}
\end{center}
\caption{Integrated luminosity needed to reach 5$\sigma$ significance in the $Z'\rightarrow \mu\mu$ channel.
lines are the results of interpolations between the points.}
\label{zprime}
\end{figure}

The dimuon decay is a golden channel for $Z'$ discovery. 
As for the RS graviton decay in dilepton, also in the case of $Z'$ the dominant background arises from the Drell-Yan
  lepton pair production, whereas contributions
from  $t\overline{t}$ and from the vector boson pair production (ZZ, WZ, WW) are significantly smaller and are highly
suppressed by selection cuts. 
The momentum resolution of the detector plays a key
role in separating the signal from the background. New reconstruction algorithms have been developed
to increase the lepton reconstruction efficiency. In particular for very high-p$_T$ muons, the track fitting in 
the tracker and in the muon system are optimised to detect and correct effects of their energy loss.
Results of a study~\cite{tdr}  
 obtained with  full simulation and reconstruction of signal and background is shown
in Fig.~\ref{zprime}, where the discovery potential of $Z'$ at CMS in the dimuon channel  is shown for six different model
predictions.

\section{Conclusions}
The initial phase of running will be crucial both for ATLAS and CMS: 
the detectors have to be understood and calibrated and the 
SM processes  have to be measured. 
After   this huge starting 
program has been done, exciting searches for new physics can be performed. 
The LHC collaborations are getting 
ready for this fascinating period, by validating the software 
and preparing data analysis while installing and commissioning the detector. 
In several   discovery  channels considered up to now, it has been demonstrated that, also
including systematic uncertainties,   signals of new physics could already
manifest with an integrated luminosity lower than 1 fb$^{-1}$.
The analysis effort in the next few months before the start of LHC will be devoted 
to further develop robust selection criteria and analysis methods to avoid biases due to
 the large systematic  uncertainties that we will have at the beginning.

\end{document}